\documentclass[
    ,final            ]
  {aipproc}

\layoutstyle{6x9}

\begin{document}

\title{NLO forward jet vertex}

\classification{12.38.-t,12.38.Cy}
\keywords      {perturbative QCD, BFKL approach, jets, LHC}

\author{F.~Caporale}{
  address={Dipartimento di Fisica, Universit\`a della Calabria, and Istituto 
Nazionale di Fisica Nucleare, Gruppo collegato di
Cosenza, I-87036 Arcavacata di Rende, Cosenza, Italy}
}

\author{D.Yu.~Ivanov}{
   address={Sobolev Institute of Mathematics and
Novosibirsk State University, 630090 Novosibirsk, Russia}
}

\author{B.~Murdaca}{
  address={Dipartimento di Fisica, Universit\`a della Calabria, and Istituto 
Nazionale di Fisica Nucleare, Gruppo collegato di
Cosenza, I-87036 Arcavacata di Rende, Cosenza, Italy}
}

\author{A.~Papa}{
  address={Dipartimento di Fisica, Universit\`a della Calabria, and Istituto 
Nazionale di Fisica Nucleare, Gruppo collegato di
Cosenza, I-87036 Arcavacata di Rende, Cosenza, Italy}
}
\author{A.~Perri}{
  address={Dipartimento di Fisica, Universit\`a della Calabria, and Istituto 
Nazionale di Fisica Nucleare, Gruppo collegato di
Cosenza, I-87036 Arcavacata di Rende, Cosenza, Italy}
}

\begin{abstract}
We calculate in the BFKL approach the jet vertex relevant for the production 
of Mueller-Navelet jets in proton-proton collisions. We consider both cases 
of incoming quark and gluon and show explicitly that all infrared divergences 
cancel when renormalized parton densities are considered. Finally we compare 
our expression for the vertex with a previous calculation~\cite{Bartels(2003)}.
\end{abstract}

\maketitle

\section{Introduction}

The large center-of-mass energy of hadron colliders like the Tevatron and the 
Large Hadron Collider (LHC) is not only useful for the production of possible 
new heavy particles, but also allows to investigate the high-energy regime of 
QCD. An especially interesting situation is the 
production of Mueller-Navelet jets~\cite{Mueller:1986ey}. This inclusive 
process $p\left( p_1\right)+p\left( p_2\right) \to  J_1\left( k_{J_1}\right) 
+J_2\left( k_{J_2}\right)+X$ was proposed to study the Regge limit of 
perturbative QCD in proton-proton collision.  
\\
Since the two jets produced have transverse momenta much larger than the QCD 
scale, $\vec k_{J_1}^2\sim \vec k_{J_2}^2
\gg \Lambda_{\rm QCD}^2$, it is possible to use perturbative QCD. Moreover, 
they are separated by a large interval of rapidity, $\Delta y\gg 1$, which 
means large center of mass energy $\sqrt s$ of the proton collisions, 
$s=2p_1\cdot p_2 \gg \vec k_{J \, 1,2}^2$.
Since large logarithms of the energy compensate the small QCD coupling, they 
must be resummed to all orders of perturbation theory.
The BFKL approach~\cite{BFKL} is the most suitable framework for the
theoretical description of the high-energy limit of hard or semi-hard
processes. It provides indeed a systematic way to perform the resummation of 
the energy logarithms, both in the leading logarithmic approximation (LLA), 
which means resummation of all terms $(\alpha_s\ln(s))^n$, and in the
next-to-leading logarithmic approximation (NLA), which means resummation of
all terms $\alpha_s(\alpha_s\ln(s))^n$. 
In QCD collinear factorization the cross section of the process reads
\[
\frac{d\sigma}{dJ_1 dJ_2}
=\sum_{i,j=q,\bar q,g}\int\limits^1_0
\int\limits^1_0 dx_1dx_2 f_i(x_1,\mu_F) f_j(x_2,\mu_F)
\frac{d\hat \sigma_{i,j}(x_1 x_2 s,\mu_F)}{dJ_1 dJ_2}\;,
\]
with $dJ_{1,2}=dx_{J_{1,2}}d^{D-2}k_{J_{1,2}}$ and the $i,j$ indices specify 
parton types (quarks $q$, antiquarks $\bar q$ or gluon $g$); $f_i(x,\mu_F)$ 
denotes the initial proton parton density function (PDF), the longitudinal
fractions of the partons involved in the hard subprocess are $x_{1,2}$,
while $\mu_F$ is the factorization scale and $d\hat
\sigma_{i,j}(x_1 x_2 s,\mu_F)$ 
is the partonic cross section for the production of jets, $\hat s=x_1 x_2 s$ 
being the energy of the parton-parton collision.
In the BFKL approach the resummed cross section of the hard subprocess reads
\[
\frac{d\hat \sigma}{dJ_1 dJ_2}=
\frac{1}{(2\pi)^{D-2}}\!\int\frac{d^{D-2} q_1}
{\vec q_1^{\,\, 2}}\frac{d\Phi_{J_1}(\vec q_1,s_0)}{dJ_1}\!\int
\frac{d^{D-2} q_2}{\vec q_2^{\,\,2}}\frac{d \Phi_{J_2}(-\vec q_2,s_0)}
{dJ_2}
\]
\vspace{-0.3cm}
$$\times\!\!\int\limits^{\delta +i\infty}_{\delta
-i\infty}\frac{d\omega}{2\pi i}\left(\frac{\hat s}{s_0}\right)^\omega
\!G_\omega (\vec q_1, \vec q_2)\, ,
$$
that is a convolution of the jet impact factors $d\Phi_i/dJ_i$
with the Green's function $G_\omega$, process-independent and determined 
through the BFKL equation.
\\
The aim of this work is the recalculation of next-to-leading order (NLO) jet 
vertex, first found by Bartels \emph{et al.}~\cite{Bartels(2003)}, in order to 
have an independent check of their results.\\
The starting point of the calculation are the impact factors for colliding 
partons~\cite{fading,fadinq,Cia,Ciafaloni:1998kx}, which describe the 
totally inclusive transition of a parton into a group of partons (one
or two in the NLO). 
In order to select the parton(s) in the final state that will generate the jet,
we ``open'' one of the integrations over the partonic phase space: 
for the one-parton case, we introduce a function $S_J^{(2)}$ that identifies 
the jet momentum with that of the parton; for the two-parton state, we 
introduce a function $S_J^{(3)}$ that identifies the jet momentum with the 
momentum of one of the two partons or with the sum of the momenta of two 
partons. In the calculation of the jet vertex, the infrared divergences related
with soft emission will cancel in the sum with virtual corrections. The 
remaining infrared divergences are taken care of by the PDFs' renormalization.
The collinear counterterms appear due to the replacement of the bare PDFs by
the renormalized physical quantities obeying DGLAP evolution equations. 
Ultraviolet divergences are removed by the counterterm related with QCD charge 
renormalization.
The LO jet impact reads:
\begin{equation}
\frac{d\Phi^{(0)}_J(\vec q\,)}{dJ}=\Phi^{(0)}_q \int_0^1 dx
\left(\frac{C_A}{C_F}f_g(x)+\sum_{a=q, \bar q} f_{a}(x)\right)S_J^{(2)}
(\vec q;x)\;,
\label{lo}
\end{equation}
where  $\Phi^{(0)}_q=g^2\frac{\sqrt{N_c^2-1}}{2N_c}$ is the quark impact 
factor at the Born level, $\vec q$ is the Reggeon momentum, $S_J^{(2)}$ is the 
selection function previously defined, and $f_g$ and $f_{a}$ the gluon and 
quark PDFs, respectively. Substituting in Eq.~(\ref{lo}) the bare QCD coupling 
and bare PDFs by the renormalized ones (in the $\overline{\rm MS}$ 
scheme), we obtain the following expressions
$$
\frac{d\Phi_J(\vec q\,)|_{\rm{charge \ c.t.}}}{dJ}
=   \frac{\alpha_s}{2\pi}\left(\frac{1}{\hat \varepsilon}+\ln\frac{\mu_R^2}
{\mu^2}\right)\left( \frac{11C_A}{6}-\frac{N_F}{3} \right)\, \Phi^{(0)}_q
$$
\vspace{-0.3cm}
\begin{equation}
\label{charge.count.t}
\times\int_0^1 dx\left(\frac{C_A}{C_F}f_g(x)+\sum_{a=q, \bar q} f_{a}(x)
\right) S_J^{(2)}(\vec q;x)
\label{ccharge}
\end{equation}
\begin{equation}
\frac{d\Phi_J(\vec q\,)|_{\rm{collinear \ c.t.}}}{dJ}
= - \frac{\alpha_s}{2\pi}\left(\frac{1}{\hat \varepsilon}+\ln\frac{\mu_F^2}
{\mu^2}\right)\Phi^{(0)}_q
\int\limits^1_{0} \, d\beta \,  \int\limits^1_{0} \,dx\,
S_J^{(2)}(\vec q\,;\beta x)
\label{cpdf}
\end{equation}
\vspace{-0.3cm}
$$
\times\left[ \sum_{a=q,\bar q}\left( P_{qq}(\beta)f_{a}\left(x\right)
+  P_{qg}(\beta) f_g\left(x\right) \right)
+\frac{C_A}{C_F}\left( P_{gg}(\beta) f_g\left(x\right)
+  P_{gq}(\beta)\sum_{a=q,\bar q}f_{a}\left(x\right)\right)
\right]\;,
$$
for the charge renormalization and the collinear counterterms, respectively
(here $\mu$ is the scale introduced by the dimensional regularization). \\
Now we have all the necessary ingredients to perform our calculation of the NLO
corrections to the jet impact factor. 
We will consider separately the subprocesses initiated by the quark and the
gluon PDFs and denote
\begin{equation}
V=V_q+V_g \, \ \ \ {\rm with} \ \ \  
\frac{d\Phi^{(1)}_J(\vec q\,)}
{dJ}\, \equiv \, \frac{\alpha_s}{2\pi}\,  \Phi^{(0)}_q \, V (\vec q\,)\; .
\end{equation}

\section{NLO jet impact factor}

In the quark case, virtual corrections are the same as in the case of the 
inclusive quark impact factor~\cite{fading,fadinq,Cia}:
\begin{equation}
V_q^{(V)}(\vec q\,)=-\frac{\Gamma[1-\varepsilon]}{\varepsilon \,
(4\pi)^\varepsilon}\frac{\Gamma^2(1+\varepsilon)}{\Gamma(1+2\varepsilon)}
 \int_0^1 dx \sum_{a=q, \bar q}
f_{a}(x) \, S_J^{(2)}(\vec q\,;x)
\label{qvirt}
\end{equation}
\vspace{-0.3cm}
$$
 \times \left[C_F\left(\frac{2}{\varepsilon}-3\right)-\frac{N_F}{3}
+C_A\left(\ln\frac{s_0}{\vec q^{\,\,2}}+\frac{11}{6}\right)
 \right] + {\rm finite \ terms} \, .
$$
For the incoming quark case, real corrections originate from the quark-gluon
production process. If we denote the transverse momentum of the gluon by $k$ 
and its longitudinal fraction by $\beta x$, then the real contribution has the 
form
$$
V^{(R)}_q\left( \vec q\right)=\int_0^1 dx\sum_{a=q, \bar q} f_{a}(x)
\left\lbrace\frac{\Gamma[1-\varepsilon]}{\varepsilon \,(4\pi)^\varepsilon}
\frac{\Gamma^2(1+\varepsilon)}{\Gamma(1+2\varepsilon)} 
\left[ C_F\left( \frac{2}{\varepsilon}-3\right)\,
S_J^{(2)}(\vec q\,;x)\,\right.\right.
$$
\vspace{-0.3cm}
\begin{equation}
\label{CF_quark}
\left.+ \int_0^1 d\beta \, \left(\,P_{qq}\left( \beta \right)
+\frac{C_A}{C_F}P_{gq}\left( \beta \right)\right)S_J^{(2)}\left( \vec q; x
\beta\right)\right] 
+\frac{C_A}{(4\pi)^{\varepsilon}}\int \frac{d^{D-2} k}
{\pi ^{1+\varepsilon}}
\label{qreal}
\end{equation}
\vspace{-0.3cm}
$$\left.
\times\frac{\vec q^{\,\, 2}}{\vec k^{\,\, 2}\left( \vec q-\vec k\right)^2}
\ln\frac{s_0}{ \left(|\vec k| + |\vec q-\vec k |\right)^2}\, S_J^{(2)}
\left( \vec q-\vec k; x\right)\right\rbrace+ {\rm finite \ terms}\ .
$$
Also in the gluon case the virtual corrections are the same as for the 
inclusive gluon impact factor~\cite{fading,Cia}:
\begin{equation}
V_g^{(V)}(\vec q)=-
\frac{\Gamma[1-\varepsilon]}{\varepsilon\,  (4\pi)^\varepsilon}
\frac{\Gamma^2(1+\varepsilon)}{\Gamma(1+2\varepsilon)}\,
\int_0^1 dx \, \frac{C_A}{C_F}\, f_g(x) \, S_J^{(2)}(\vec q\,;x) 
\label{gvirt}
\end{equation}
\vspace{-0.3cm}
$$
\times\left[\, C_A\ln\left(\frac{s_0}{\vec q^{\: 2}}\right)
+C_A\left( \frac{2}{\varepsilon}-\frac{11}{6}\right) + \frac{N_F}{3} \right]
+ {\rm finite \ terms} \; .
$$
In the NLO gluon impact factor real corrections come from the production
of quark-antiquark state and a gluon-gluon 
state~\cite{fading,Cia,Ciafaloni:1998kx}. We find
$$
V_g^{(R)}(\vec q\,)=  \frac{\Gamma[1-\varepsilon]}{\varepsilon\,  
(4\pi)^\varepsilon} \frac{\Gamma^2(1+\varepsilon)}{\Gamma(1+2\varepsilon)}
\,\int_0^1 dx  \, f_g(x) \, \left\lbrace \frac{C_A}{C_F}\left(\frac{N_F}{3}
+\frac{2C_A}{\varepsilon}-\frac{11}{6}C_A\right)\,\right.
$$
$$
\times S_J^{(2)}(\vec q;x)+\,
\int_0^1 d\beta \, \left[2N_F P_{qg}(\beta)+2C_A\frac{C_A}{C_F}\left(P(\beta)
+\frac{(1-\beta)P(1-\beta)}{(1-\beta)_+}\right)\right]\,
$$
$$
\left.\times S_J^{(2)}(\vec q;x\beta) \right\rbrace
+\, \frac{C_A }
{(4\pi)^{\varepsilon}} \int_0^1 dx \, \frac{C_A}{C_F} \, f_g(x)
\int \frac{d^{D-2} k}{\pi^{1+\varepsilon}}
\frac{\vec q^{\: 2}}{\vec k^{ 2}(\vec k-\vec q\,)^2}
\ln \frac{s_0}{(|\vec k|+|\vec q-\vec k|)^2} 
$$
\vspace{-0.2cm}
\begin{equation}
\times\, S^{(2)}_J(\vec q-\vec k;x)+\,  {\rm finite \ terms}\; .
\label{greal}
\end{equation}
To conclude, we collect the contributions given in 
Eqs.~(\ref{ccharge}),~(\ref{cpdf}),~(\ref{qvirt})-(\ref{greal}), and we note 
that we are left with two divergences: the last terms of~(\ref{qreal}) and 
of~(\ref{greal}). It easy to see that the convolution of the jet vertex with 
BFKL Green's function, required for the calculation of the jet cross section,
will give a divergence-free result. 
\\
More details about this calculation can be found in Ref.~\cite{Caporale}.

\section{Summary}

We have recalculated the jet production vertex, first found by Bartels 
\emph{et al.}~\cite{Bartels(2003)}, in a more direct way, starting from the 
known general expression of NLO impact factors, given in Ref.~\cite{FF98},
and applied to the case of parton impact factors. Nevertheless, in many 
technical steps we followed closely the derivation of 
Refs.~\cite{Bartels(2003)}. In our approach the energy scale $s_0$ is not 
fixed. Performing the transition (see~\cite{Fadin:1998sg}) from the standard 
BFKL scheme with arbitrary energy scale $s_0$ to the one used 
in~\cite{Bartels(2003)}, where the energy scale depends on the Reggeon 
momentum, we can see a complete agreement with 
Refs.~\cite{Bartels(2003)}.\\
The jet vertex discussed in this paper is an essential ingredient also for the 
study of the forward jet production in deep inelastic scattering in the NLA.

\end{document}